# Extended Unary Coding

Subhash Kak[1]

**Abstract**
Extended variants of the recently introduced spread unary coding are described. These schemes, in which the length of the code word is fixed, allow representation of approximately $n^2$ numbers for *n* bits, rather than the n numbers of the standard unary coding. In the first of two proposed schemes the spread increases, whereas in the second scheme the spread remains constant.

1. **Introduction**

Recently a variant of the unary code called the spread unary code (SU code) was introduced [1]. The motivation for this was to give a mathematical form to the way unary coding works in biology, where the number marker is not necessarily a single neuron [2],[3]. The SU code is more robust than standard unary coding and thus suitable for practical applications.

Let the size of the code word be *n* bits. In the unary code, the number *i* is simply a concatenation of *i* 1s. To frame the number, we append a 0 to the left (or to the right if one uses a different convention). Thus 5 is 011111. In its spatial form, the number will be denoted by the leftmost 1, so that 5 becomes 010000. In the spread unary code, we replace each 1 in the spatial form with *k* 1s (and this may require increasing the size of the code word). Thus the number *i* will be represented by *k* 1s followed by *i-1* 0s:

*i: 1 (*k times*)0 (*i-1 times*)* (1)

Thus 5 now becomes (say for *k*=3) 1110000. The minimum size of the code word is now $k+i-1$ which in this case is 7 and larger than the previous case where it was 6.

Table 1. Spread unary code for *k*=2
for numbers 1 through 10

| *i* | Spread unary code |
|---|---|
| 0 | 00000000000 |
| 1 | 00000000011 |
| 2 | 00000000110 |
| 3 | 00000001100 |
| 4 | 00000011000 |
| 5 | 00000110000 |
| 6 | 00001100000 |
| 7 | 00011000000 |
| 8 | 00110000000 |
| 9 | 01100000000 |
| 10 | 11000000000 |

Note that the value of the number in Table 1 is according to the location of the rightmost 1. Although it has the virtue of being distributed, its downside is that it is sparse like the other forms of unary code. For example, the numbers 0 through 10 have a total of 30 1s and 120 0s. We would like to have a scheme which is more efficient.

---

[1] Oklahoma State University, Stillwater, OK 74078

In this paper we present efficient scheme of number representation using an extended variant of spread unary. We call this new scheme EU (extended unary). We will show that for *n* bits it makes it possible to represent number from 0 to $(n-k)^2-1$, where $k \geq 2$. For example rather than using 100 unary bits, we represent the numbers 1 through 99 using only 12 bits (given $k=2$).

## 2. Extended Unary Coding

The objective is to find an efficient representation that utilizes the zeros that lie around the 1s that are sparse since k is much less than 1. This utilization of the 0s to pack in additional information can be done in many different ways. Here we describe two methods in the first of which the value of k increases, and in the second it remains constant.

*First Code. Increasing k* (**EU-ik**)
Here the count goes on successively employing increasing larger k, until $k = n$. Example of this for $n = 3$ is:

    0: 000
    1: 001 (*first cycle*)
    2: 010
    3: 100
    4: 011 (*second cycle*)
    5: 110
    6: 111 (*third cycle*)

So with $n = 3$, one could count from 0 to 6. Generalizing this construction, we have the following result.

**Theorem 1.** The total count for n bits in the increasing-k (EU-ik) scheme is 0 through $n(n+1)/2$.

*Proof.* The first cycle will count to *n*, the second to *n-1*, and so on. The total, therefore, is
$n + (n-1) + (n-2) + ... + (n-n+1) = n^2 - (1+2+3+...+n-1) = n(n+1)/2$.

*Second Code. Fixed k* (**EU-fk**)
In this $k$ is fixed and after the first $k+1$ digits the additional digits are marked by a 1 that is separated from the basic set of $k$ 1s, which requires that $n > k+1$. The separation is at first 1 unit and then it is successively increased. An example of this for $n=7$ and $k=3$ is:

    0: 0000000
    1: 0000111 (*beginning of first cycle*)
    2: 0001110
    3: 0011100
    4: 0111000
    5: 1110000
    6: 0010111 (*second cycle marked by a 1 separated by one 0*)
    7: 0101110
    8: 1011100
    9: 0111001 (*the marker has cyclically shifted to the right*)
    10: 1110010 (*end of second cycle*)

11: 0100111 (*third cycle marked by a 1 separated by two 0s*)
12: 1001110
13: 0011101 (*the marker cyclically shifts to the right*)
14: 0111010
15: 1110100

**Theorem 2.** In the fixed $k$ method for $n$ bits ($n > k+1$) where the additional cycles are marked by 1 at distance of 1 and more in succession, the total count is 0 through $(n-k)^2 - 1$.

*Proof:* The condition $n > k+1$ is essential because the counting scheme requires a gap of one zero between the cycle marker and the bits representing the number. The count is $(n-k+1)$ in the first cycle, and each of the subsequent cycles (unless it is zero which would be the case if $n = k+1$). The total number of cycles possible is $(n-k-1)$. Therefore the total count is $(n-k+1)(n-k-1) = (n-k)^2 - 1$.

*Example.* Let $n = 11$, and $k = 2$ and let us use the second efficient spread unary code. This code of 11 bits will be able to count from 0 through 80.

Table 2. EU-fK unary code for $k = 2$ for numbers 1 through 40

| i | Spread unary code | i | Spread unary code |
|---|---|---|---|
| 0 | 00000000000 | | |
| 1 | 00000000011 | 11 | 00000001011 |
| 2 | 00000000110 | 12 | 00000010110 |
| 3 | 00000001100 | 13 | 00000101100 |
| 4 | 00000011000 | 14 | 00001011000 |
| 5 | 00000110000 | 15 | 00010110000 |
| 6 | 00001100000 | 16 | 00101100000 |
| 7 | 00011000000 | 17 | 01011000000 |
| 8 | 00110000000 | 18 | 10110000000 |
| 9 | 01100000000 | 19 | 01100000001 |
| 10 | 11000000000 | 20 | 11000000010 |
| 21 | 00000010011 | 31 | 00000100011 |
| 22 | 00000100110 | 32 | 00001000110 |
| 23 | 00001001100 | 33 | 00010001100 |
| 24 | 00010011000 | 34 | 00100011000 |
| 25 | 00100110000 | 35 | 01000110000 |
| 26 | 01001100000 | 36 | 10001100000 |
| 27 | 00011000000 | 37 | 00011000001 |
| 28 | 00110000001 | 38 | 00110000010 |
| 29 | 01100000010 | 29 | 01100000100 |
| 30 | 11000000100 | 40 | 11000001000 |

Table 3. EU-fK for $k=2$, $n=11$ for some random numbers

| i | EU-fk code |
|---|---|
| 43 | 00100001100 |
| 46 | 00001100001 |
| 58 | 00110001000 |
| 77 | 00011010000 |
| 80 | 11010000000 |

### 3. Algorithms for coding and decoding

*First Code (EU-ik):*
Let the number to be code be in the range (0, N). Given the condition of Theorem 1, it follows that the relationship between N and n is given by $2N \leq n^2+n$.

| N | n |
|---|---|
| 10 | 4 |
| 15 | 5 |
| 21 | 6 |
| 28 | 7 |
| 36 | 8 |
| 45 | 9 |
| 55 | 10 |
| 66 | 11 |
| 78 | 12 |
| 210 | 20 |
| 1275 | 50 |

Let's suppose we wish to represent the number 48 for *n*=11. In this case, the number of 1s in the code will be 6 because with n=11, the count will be 11, 10, 9, 8, 7 (a total of 45) for the spread codes with 1 through 5 1s. Therefore 46: 00000111111; 47: 00001111110; and 48: 00011111100.

*Second Code (EU-fk):*
Here the condition from Theorem 2 is $N \leq (n-k)^2-1$.

The following table gives us the relationship:

| N | n | k |
|---|---|---|
| 15 | 6 | 2 |
| 24 | 7 | 2 |
| 35 | 8 | 2 |
| 48 | 9 | 2 |
| 63 | 10 | 2 |
| 80 | 11 | 2 |
| 99 | 12 | 2 |
| 120 | 13 | 2 |
| | | |
| 8 | 6 | 3 |
| 15 | 7 | 3 |
| 24 | 8 | 3 |
| 35 | 9 | 3 |
| 48 | 10 | 3 |
| 63 | 11 | 3 |
| | | |
| 15 | 8 | 4 |
| 24 | 9 | 4 |
| 35 | 10 | 4 |
| 48 | 11 | 4 |
| 63 | 12 | 4 |
| 15 | 9 | 5 |
| 24 | 10 | 5 |
| 35 | 11 | 5 |
| 48 | 12 | 5 |
| 63 | 13 | 5 |

The capacity of a choice depends on $n-k$. Thus to get a count of up to 63, we can choose $(n,k)$ to be either (10,2), or (11,3), or (12,4), or (13,5), and so on.

**Conclusions**

Extended variants of the recently introduced spread unary coding are described. These schemes, in which the length of the code word is fixed, allow representation of approximately $n^2$ numbers for $n$ bits, rather than the $n$ numbers of the standard unary coding. In the first scheme the spread increases, whereas in the second scheme the spread remains constant. We should add that other variants of the spread unary code can be derived.

Since the unary code has found uses in instantaneously trained neural networks [4]-[6], it is worthwhile to ask if the extended unary code will also be of value in such applications.

**References**

1. S. Kak, Spread unary coding. 2014. arXiv:1412.6122
2. I.R. Fiete, R.H. Hahnloser, M.S. Fee, and H.S. Seung, Temporal sparseness of the premotor drive is important for rapid learning in a neural network model of birdsong. J Neurophysiol., 92(4), pp. 2274–2282, 2004.
3. I.R. Fiete and H.S. Seung, Neural network models of birdsong production, learning, and coding. New Encyclopedia of Neuroscience. Eds. L. Squire, T. Albright, F. Bloom, F. Gage, and N. Spitzer. Elsevier, 2007.
4. K.-W. Tang and S. Kak, A new corner classification approach to neural network training. Circuits, Systems, and Signal Processing 17: 459-469, 1998.
5. S. Kak, Faster web search and prediction using instantaneously trained neural networks. IEEE Intelligent Systems 14: 79-82, November/December 1999.
6. S. Kak, A class of instantaneously trained neural networks. Information Sciences 148: 97-102, 2002.